\documentstyle[11pt,aas2pp4,psfig]{article}
\lefthead{CANZIAN}
\righthead{EXTENSIVE SPIRAL STRUCTURE}

\begin{document}

\title{EXTENSIVE SPIRAL STRUCTURE AND COROTATION RESONANCE}

\author{Blaise Canzian\altaffilmark{1}}
\affil{Universities Space Research Association,\\
U.S. Naval Observatory, and\\
Space Telescope Science Institute(email:  blaise@nofs.navy.mil)}
\altaffiltext{1}{Postal address:  U.S. Naval Observatory, PO Box 1149,
Flagstaff, AZ 86002}

\begin{abstract}
Spiral density wave theories demand that grand design spiral structure
be bounded, at most, between the inner and outer Lindblad resonances of
the spiral pattern.  The corotation resonance lies between the outer
and inner Lindblad resonances.  The locations of the resonances are at
radii whose ratios to each other are rather independent of the shape of
the rotation curve.  The measured ratio of outer to inner extent of
spiral structure for a given spiral galaxy can be compared to the
standard ratio of corotation to inner Lindblad resonance radius.  In
the case that the measured ratio far exceeds the standard ratio, it is
likely that the corotation resonance is within the bright optical
disk.  Studying such galaxies can teach us how the action of resonances
sculpts the appearance of spiral disks.  This paper reports
observations of 140 disk galaxies, leading to resonance ratio tests for
109 qualified spirals.  It lists candidates that have a good chance of
having the corotation resonance radius within the bright optical disk.
\end{abstract}

\keywords{galaxies: kinematics and dynamics; galaxies: spiral;
galaxies: structure}

\section{Introduction}

Disk galaxies exhibit many designs:  bars, rings, and spiral
structure.  The structure and placement of rings and bars are widely
believed, on numerical and analytical grounds, to be governed by
resonance.  A resonance is a rational relationship between two
frequencies.  Inner and outer rings form at radii that are resonant
with the angular frequency of the bar pattern (Schwarz
\markcite{Schw81}1981; Buta \markcite{Buta86}1986; Buta \& Crocker
\markcite{Buta91}1991).  A strong bar ends near its corotation
resonance (Contopoulos \markcite{Cont80}1980), referred to hereafter as
``CR''.

Many disk galaxies show spiral patterns, some of which are beautiful,
two-armed spirals.  What effect do resonances have on the appearance of
spiral patterns?  First consider barred spirals.  It was once widely
believed that the bar was capable of ``driving'' spiral structure
(e.g., van Albada \& Roberts \markcite{vanA81}1981) or of providing a
source of disturbances that might be swing amplified into spiral
structure (Toomre \markcite{Toom77}1977).  Then the bar and the spiral
patterns would be resonantly related.  But Sellwood \& Sparke
(\markcite{Sell88}1988) have suggested that spiral patterns need not be
resonantly related to bars.  Kinematic data for few galaxies have been
analyzed to see if the resonant link exists between bars and spirals.
Bar and spiral are part of the same pattern in both NGC1365 and
NGC4321, for example, but the transition from bar to spiral is at the
CR in NGC1365 (Lindblad, Lindblad, \& Athanassoula
\markcite{Lind96}1996) while it is at the inner 4:1 resonance in
NGC4321 (Elmegreen, Elmegreen, \& Montenegro \markcite{Elme92}1992;
Sempere et al. \markcite{Semp95}1995; Canzian \& Allen
\markcite{Canz97}1997).  Apart from these two galaxies, there are
differences between strongly and weakly barred spirals to consider, as
well as differences between flat and exponential bars.

Spiral patterns exist in disks without bars, too.  Spiral density waves
are constrained to exist between the inner and outer Lindblad
resonances, hereafter ``ILR'' and ``OLR''.  Aside from this constraint,
the role of resonances in shaping pure spiral morphology is uncertain.

Many signatures of resonance structures have been suggested and used
with varying success.  Dust lanes may jump to the other sides of spiral
arms across the CR (Roberts \markcite{Robe69}1969).  Star formation may
be less efficient within spiral arms near CR (seen in NGC0628 and
NGC3992 by Cepa \& Beckman \markcite{Cepa90}1990 and in M51 by Knapen
et al.  \markcite{Knap92}1992 but not in NGC4321 by Knapen et al.
\markcite{Knap96}1996).  Enhanced star formation may occur in the
interarm region at the CR, and at the inner 4:1 resonance there may be
brightness minima in the arms and spurs should erupt in the interarm
region (Elmegreen et al.  \markcite{Elme89}1989,
\markcite{Elm92}1992).  Not all of the above prescriptions work in all
cases.

Tests of the above prescriptions can follow only when clear
determinations of resonance locations have been made for many spirals.
Confirmation of resonance locations requires kinematic data.  Such data
derive from emission line mapping, which is difficult and time
consuming.  It would be more rewarding to study galaxies for which
interesting resonance locations were known beforehand to be at easily
studied radii in the bright disk.  The CR is one of the most
interesting resonances to study.

This paper presents a test to identify galaxies whose CR is favorably
placed for study (within the bright disk).  The test requires only
moderately deep imaging in typical seeing.  No rotation curve data are
necessary.  The test is not completely determinate:  galaxies with CR
within the bright disk can be missed.  This is the price paid for
operating in ignorance of the rotation curve shape and spiral pattern
speed.  However, the test is reliable:  when a suitable galaxy is
identified, there is a good chance that its CR is, indeed, within the
bright disk.

The basis of the test to identify galaxies with well-placed CR is
presented in \S\ref{sec_extensive}.  The section also investigates how
much effect the shape of the rotation curve and the spiral pattern
speed have on the outcome of the test.  \S\ref{sec_observations}
outlines the observations.  The analysis of the imaging is described in
\S\ref{sec_analysis}.  This includes not only a description of the
measurements made but also some surface photometry to estimate how deep
the imaging reaches.  The results are presented in \S\ref{sec_discussion},
where promising galaxies are described and where the different
characteristics of barred and pure spirals are discussed.

\section{Extensive spiral structure}\label{sec_extensive}

Spiral structure is constrained to be between the ILR and OLR of the
spiral pattern.  The CR is between the ILR and OLR.  It is not known
how often spiral structure remains relatively bright, or even optically
detectable, at radii approaching the OLR.  Spiral structure is probably
detectable at the CR in most galaxies.  We seek galaxies for which the
spiral structure is not merely detectable, but is genuinely bright, at
the CR.  By implication, the spiral structure would then stretch most
of the way between the ILR and OLR.

To test for placement of the CR within the bright disk, first measure
the inner and outer extent of the spiral structure from deep imaging in
typical seeing.  When the measured ratio of the outer to inner extent
exceeds the ``typical'' ratio of CR to ILR radius (described in
\S\ref{sec_flatrc}), the spiral structure appears to have needed more
space than is available just between the ILR and CR.  Therefore, the CR
is in the optically detectable part of the disk.  The larger the
measured ratio, the deeper into the bright part of the disk the CR
might be.

\subsection{Flat rotation curve}\label{sec_flatrc}

\begin{figure}
\figurenum{1}
\plotone{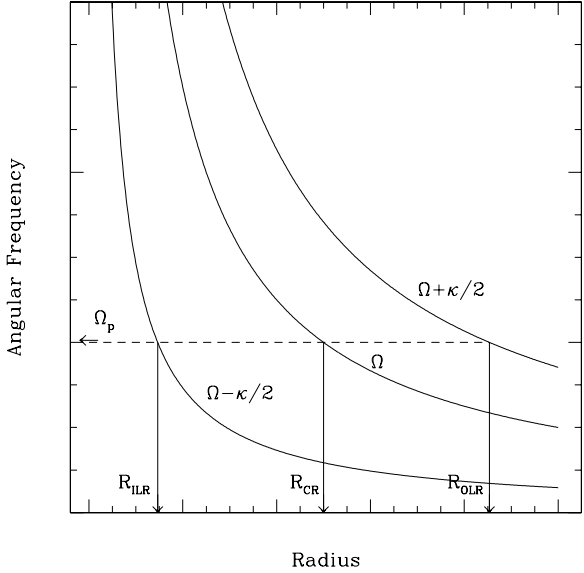}
\caption{Schematic showing angular frequencies (in
arbitrary units) for a flat rotation curve.  Angular rotation frequency
is $\Omega$ and the epicyclic frequency is $\kappa$.  The pattern
``speed'' is $\Omega_p$.  Corotation resonance (CR) and inner (ILR) and
outer (OLR) Lindblad resonances are marked.\label{fig_flatfreq}}
\end{figure}

Consider a flat
rotation curve.  (More realistic shapes will be considered in
\S\ref{sec_shape}.)  At the CR, disk material
has an orbital frequency equal to the angular frequency of the
spiral pattern:  the stars and pattern corotate.  That is,
\begin{equation}
\Omega_p = V/R_{\rm CR}
\end{equation}
at the CR, where $\Omega_p$ is the pattern
angular frequency, $V$ is the circular rotation speed, and $R_{\rm CR}$
is the corotation resonance radius.  At the Lindblad resonances, the
material experiences the entire pattern exactly once each orbit.  At
the ILR, the material sweeps through the pattern,
while at the OLR, the pattern catches up with and
passes through the material.  Consider a two-armed spiral pattern.
At the Lindblad resonances,
\begin{equation}
\Omega_p = \Omega\pm\kappa/m,
\end{equation}
where $\kappa$ is the epicyclic frequency, the $+$ sign refers to OLR,
the $-$ sign to ILR, and $m$ is the number of spiral arms.  (Only
two-armed spirals will be considered in this paper.)
Figure~\ref{fig_flatfreq} shows graphs of angular frequencies computed
for a flat rotation curve.  $\Omega - \kappa/2$ is a decreasing
function exterior to the ILR.

Let $R_1$ be the innermost and let $R_2$ be the outermost measured
radii of the spiral structure.  $R_2 \le R_{\rm CR}$ implies
\begin{equation}
(\Omega - \kappa/2)_{\rm ILR} = V/R_{\rm CR} \le V/R_2.\label{eq_a}
\end{equation}
The epicyclic frequency is (e.g., Binney \& Tremaine \markcite{Binn87}1987)
\begin{equation}
\kappa(R)\equiv \left[ 2\frac{V}{R}\left(\frac{dV}{dR} +
\frac{V}{R}\right)\right]^{1/2},
\end{equation}
and for a flat rotation curve,
\begin{equation}
\kappa=\sqrt{2}V/R.\label{eq_kappa}
\end{equation}
Equations~\ref{eq_a} and~\ref{eq_kappa} lead to
\begin{equation}
R_{\rm ILR} \ge (1-\sqrt{2}/2)R_2.\label{eq_RilrR2}
\end{equation}
Since $R_{\rm ILR}\le R_1$, equation~\ref{eq_RilrR2} implies
\begin{equation}
(1-\sqrt{2}/2)R_2/R_1 \le 1\label{eq_ilrcr}
\end{equation}
is the test of whether the observed spiral structure could fit between
the ILR and CR.  See \S\ref{sec_shape} for necessary modifications and
interpretation of equation~\ref{eq_ilrcr} when the rotation curve is
not flat.  The numerical coefficient in equation~\ref{eq_ilrcr} was
referred to in \S\ref{sec_extensive} as the ``typical'' ratio.

What is the characteristic of spiral structure bounded between the ILR
and OLR?  $R_{\rm ILR}\le R_1$ implies that
\begin{equation}
(\Omega - \kappa/2)_{R_1} \le (\Omega - \kappa/2)_{\rm ILR} =
V/R_{\rm CR}.\label{eq_R1Rilr}
\end{equation}
Equations~\ref{eq_R1Rilr} and~\ref{eq_kappa} imply that
\begin{equation}
R_{\rm CR} \le (2+\sqrt{2})R_1.\label{eq_R1Rcr}
\end{equation}
Since $R_2\le R_{\rm OLR}$,
\begin{equation}
(\Omega + \kappa/2)_{R_2} \ge (\Omega + \kappa/2)_{\rm OLR} =
V/R_{\rm CR}.\label{eq_R2Rolr}
\end{equation}
Then equations~\ref{eq_kappa} and~\ref{eq_R2Rolr} imply
\begin{equation}
(2-\sqrt{2})R_2 \le R_{\rm CR}.\label{eq_R2Rcr}
\end{equation}
Combining equations~\ref{eq_R1Rcr} and~\ref{eq_R2Rcr} leads to the condition
\begin{equation}
R_2/(3+2\sqrt{2})R_1 \le 1.\label{eq_ilrolr}
\end{equation}

\subsection{Other rotation curves}\label{sec_shape}

How does the shape of the rotation curve change the results of
\S\ref{sec_flatrc}?  Four functional forms for rotation curves will be
explored.  Rotation curves that are not strictly flat lead to numerical
coefficients somewhat different than, but still close to, those in
equations~\ref{eq_ilrcr} and~\ref{eq_ilrolr}, so small violations of
equations~\ref{eq_ilrcr} or~\ref{eq_ilrolr} are expected.

\begin{figure}
\figurenum{2}
\plotone{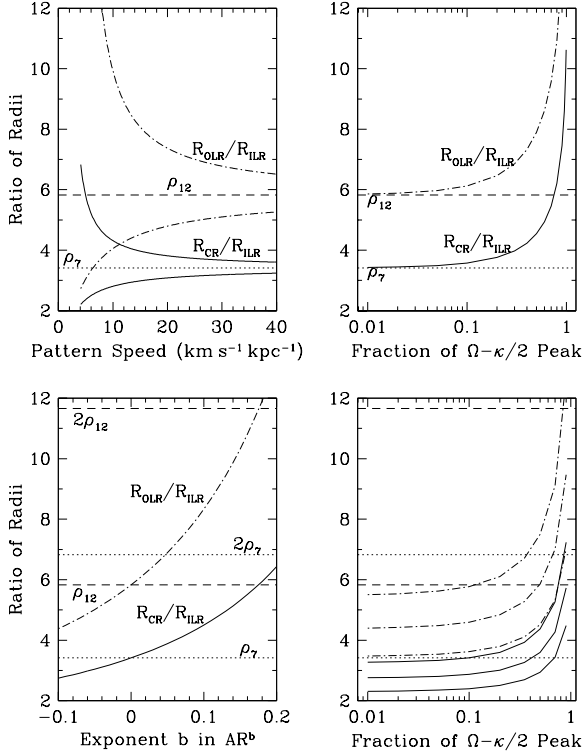}
\caption{Graphs of the resonance ratios $R_{\rm CR}/R_{\rm
ILR}$ (solid lines) and $R_{\rm OLR}/R_{\rm ILR}$ (dot-dash lines) for
four rotation curve shapes:  {\em (a)} linear rising (upper) or falling
(lower) with slope 10\% over $10\rm\,kpc$; {\em (b)}
equation~\protect\ref{eq_rcfamily} family; {\em (c)} power law; {\em
(d)} three members of the equation~\protect\ref{eq_hybrid} family with
$c=-0.01$ (upper), $-0.02$ (middle), and $-0.1$ (lower).
$\rho_7=2+\protect\sqrt{2}$, from equation~\protect\ref{eq_ilrcr}, and
$2\rho_7$ are shown by dotted lines.  $\rho_{12}=3+2\protect\sqrt{2}$,
from equation~\protect\ref{eq_ilrolr}, and $2\rho_{12}$ are shown with
dashed lines.  The pattern speed is represented in panels {\em (b)} and
{\em (d)} by its value as a fraction of the peak in $\Omega-\kappa/2$.
\label{fig_4graph}}
\end{figure}

A rotation curve that is linearly rising or falling (having functional
form $V(R)=a+bR$ for $a\neq 0$ and $b$ small) will alter the values of
the ``typical'' resonance ratios (equation~\ref{eq_ilrcr} or
equation~\ref{eq_ilrolr}).  The changes are generally small and are
independent of the value of $a$.  Figure~\ref{fig_4graph}{\em a} shows
the variation of the $R_{\rm CR}/R_{\rm ILR}$ and $R_{\rm OLR}/R_{\rm ILR}$
ratios for $b=\pm 10\%(a/10\rm\,kpc)$.  The $R_{\rm CR}/R_{\rm ILR}$ curve
is asymptotic to $\rho_7\equiv 2+\sqrt{2}$ and the $R_{\rm
OLR}/R_{ILR}$ curve is asymptotic to $\rho_{12}\equiv 3+2\sqrt{2}$,
diverging strongly only for very low pattern speeds.

Consider the following family of rotation curves:
\begin{equation}
V(R) = \frac{aR}{1+bR},\label{eq_rcfamily}
\end{equation}
where $V(R)$ is the rotation speed as a function of radius $R$ and $a$
and $b$ are free parameters.  For small $R$, $V\sim aR$ is
approximately linear, and for large $R$, $V\sim a/b$ is approximately
constant, thus approximating the shapes of many real rotation curves.

\begin{figure}
\figurenum{3}
\plotone{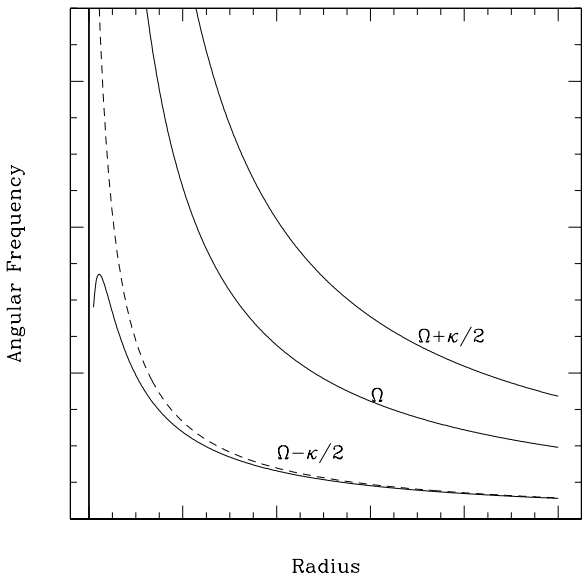}
\caption{This figure is similar to
Fig.~\protect\ref{fig_flatfreq} but displays angular frequency curves
(solid lines) for a representative member of the
equation~\protect\ref{eq_rcfamily} family.  The dashed line shows for
comparison the graph of $\Omega-\kappa/2$ for a flat rotation curve
with the same asymptotic rotation speed.  A vertical solid line is
drawn at zero radius.\label{fig_freq}}
\end{figure}

Graphs of the ratios $R_{\rm CR}/R_{\rm ILR}$ and $R_{\rm OLR}/R_{\rm
ILR}$ are shown in Figure~\ref{fig_4graph}{\em b} for the
equation~\ref{eq_rcfamily} family.  The ratios are independent of
the asymptotic rotation speed and the slope of the inner part.  The
ratios depend weakly on pattern speed until the pattern speed is a
significant fraction (over 50\%) of the value at the peak of the
$\Omega-\kappa/2$ curve.  The curvature of 
the $\Omega-\kappa/2$ curve (see Figure~\ref{fig_freq}) pulls
the ILR to a smaller radius than for a flat rotation curve.  The
smaller inner radius, used as the divisor in equation~\ref{eq_ilrcr} or
equation~\ref{eq_ilrolr}, will cause the computed value exceed unity.
For the equation~\ref{eq_rcfamily} rotation curves, the
equation~\ref{eq_ilrcr} value reaches 1.71 when the pattern speed is
75\% of the angular frequency at the peak in $\Omega-\kappa/2$.  (When
the equation~\ref{eq_ilrcr} value is 1.71, the equation~\ref{eq_ilrolr}
ratio is unity.)  The equation~\ref{eq_ilrcr} ratio equals 2 when the
pattern speed is at 86\% of the angular frequency at the peak.  If the
pattern speed exceeds the value at the peak, then there is no ILR, and
spiral structure may (in principle) extend to the center of the disk.
Then the $R_{\rm CR}/R_{\rm ILR}$ and $R_{\rm OLR}/R_{\rm ILR}$ ratios
can be arbitrarily large.

Equations~\ref{eq_ilrcr} and~\ref{eq_ilrolr} will be called the
``canonical'' tests; replacing the unit value of the inequality by 2
will be called the ``conservative'' tests.  When the conservative
equation~\ref{eq_ilrcr} test is applied, we are protecting ourselves
against the case (a ``false positive'') of a galaxy with a non-flat
rotation curve giving a false indication that its CR is within its
bright optical disk.

The power law is another family of rotation curve forms:
\begin{equation}
V(R)=aR^b.\label{eq_powlaw}
\end{equation}
Many dwarf galaxies and a moderate fraction of low surface brightness
disk galaxies have rotation curves that are well approximated by a
power law out to the farthest observed radius (de Blok, McGaugh, \& van
der Hulst \markcite{Blok96}1996; Pickering et al.
\markcite{Pick97}1997).  Current thinking on galaxy formation suggests
that rotation curves that have an inner power-law form will flatten out
at larger radius (Dalcanton, Spergel, \& Summers
\markcite{Dalc97}1997).  But it is nonetheless worth while to derive
the resonance ratio tests when the rotation curve is a power law, since
it is possible that the ILR and OLR are within the power law-like
region of the rotation curve.

Rising rotation curves are given by positive values of the exponent
$b$, with small values approximating a flat rotation curve; slowly
falling but nearly flat curves are given by negative values of the
exponent $b$.  For power law rotation curves,
\begin{equation}
R_{\rm CR}/R_{\rm ILR} = \left[ 1 - \sqrt{(b+1)/2}\right]^{1/(b-1)}.\label{eq_powlaw_ilrcr}
\end{equation}
and
\begin{equation}
R_{\rm OLR}/R_{\rm ILR} = \left[ 2\left(1 - \sqrt{(b+1)/2}\right)^2/(1-b)\right]^{1/(b-1)}.\label{eq_powlaw_ilrolr}
\end{equation}
Graphs of the variation of these ratios with exponent $b$ are shown in
Figure~\ref{fig_4graph}{\em c}.  Although the ratios are a bit smaller
than the canonical test values (equations~\ref{eq_ilrcr}
and~\ref{eq_ilrolr}) for the smallest reasonable values of $b$, the
range between the canonical test value and the conservative test value
includes most of the variation of the ratio graphs.

Consider finally the hybrid form (Elmegreen \& Elmegreen \markcite{Elme90}1990):
\begin{equation}
V(R) = aR/\left[R^b + R^{(1-c)}\right].\label{eq_hybrid}
\end{equation}
Some examples of rotation curves of this shape are shown in Fig.~4 in
Elmegreen \& Elmegreen (\markcite{Elme90}1990).  Numerical solution for
the resonance ratios led to the curves graphed in
Figure~\ref{fig_4graph}{\em d}.  In this case, too, the range between
the canonical test value and the conservative test value includes much
of the variation in the resonance ratios shown in Fig.~\ref{fig_4graph}{\em d}.

In all cases of realistic rotation curve shapes examined above, most of
the variation in the resonance ratios falls within the range between
the canonical and conservative tests derived for a flat rotation
curve.  For many rotation curves, the ratio is smaller than that for a
flat rotation curve (see, e.g., Fig.~\ref{fig_4graph}{\em d}).  The
consequence of small ratios is that some interesting candidates may not
be noticed.  Galaxies with such rotation curves might have only a
modest extent of spiral structure as measured by the canonical test,
yet the extent of spiral structure could be near the maximum allowed by
the location of the Lindblad resonances.  Such a galaxy would be missed
by the test proposed here.  Although this circumstance decreases the
efficiency of identifying promising galaxies, it also decreases the
likelihood of incorrectly selecting a galaxy that does not, in fact,
have its CR within its bright disk.

\subsection{Other caveats}\label{sec_caveats}

In a private discussion, A.~Toomre noted potential problems that might
result by assuming a unique pattern speed for the spiral structure, as
was done in \S\S\ref{sec_flatrc}-\ref{sec_shape}.  It is possible that
the spiral structure is produced from spiral density waves with a range
of pattern speeds about a well-defined characteristic pattern
speed.  There would then be both a general broadening of the observed
pattern and an apparent penetration of the Lindblad resonances of the
characteristic pattern.  This effect can be seen in Fig.~8 of Toomre
\markcite{Toom81}(1981), a classic illustration of a swing amplified
spiral density wave, where the inner parts of the spiral become tightly
wrapped and appear to penetrate the ILR.  The model was constructed
from spiral density wave packets with a Gaussian distribution of
pattern speeds.  To avoid the pitfalls of a spread of pattern speeds,
only clearly defined spiral structure was measured for this paper,
letting the seeing blur any high frequency interior structure, which
was ignored.

In contrast, spiral structure that is a mode of the disk may have a
single pattern speed, as will spiral structure produced through swing
amplification of a groove mode.  In these cases, apparent extension of
the spiral structure through Lindblad resonance should not be
apparent.

\section{Observations}\label{sec_observations}

CCD frames have been obtained for 140 galaxies.  The observations span
the period 1993 August 12 to 1997 September 7.  All observations were
made using the U.S. Naval Observatory $1\rm\,m$ telescope with any of
four CCD detectors:  either of two Texas Instruments $800^2$-pixel
CCD's (the so-called ``Banana'' and ``Flat'' chips), a Tektronix
$1024^2$-pixel CCD, or a Tektronix $2048^2$-pixel CCD.  The image scale
was $0.43\rm\,arcsec/pixel$ for the TI chips and
$0.68\rm\,arcsec/pixel$ for the Tektronix chips.  All the observations
reported here were during nights that were photometric or that had
light cirrus at worst; all frames reach comparable photometric depth.
Seeing ranged from 1.4--$3\rm\,arcsec$ full width at half maximum.
Total exposure times were 30 minutes for all regular two-armed spiral
galaxies except for the following:  NGC4321 (40 minutes), ESO437-G044
(60 minutes, but very large zenith distance), NGC3054 (40 minutes), and
NGC3433 (40 minutes).  In cases where the nucleus was inordinately
bright, two or more shorter exposures were stacked.  All exposures used
a Johnson $V$-band filter to attain large contrast in the spiral
structure with reasonable exposure time.

Two classes of galaxies make up the observing list:  physically large
spirals and grand design spirals.  Because rotation curves of spirals
are usually flat, physically large spirals have a good chance of having
spiral structure so extensive that the CR is within the bright disk.
The enormous galaxy UGC02885 (Roelfsema \& Allen \markcite{Roel85}1985;
Canzian, Allen, \& Tilanus \markcite{Canz93}1993) is the archetype of
this class.  (If rotation curves were usually linearly rising, instead
of flat, then the scale-free form of a logarithmic spiral would not
favor any particular size of spiral galaxy for this investigation.)
Data from the literature were processed to find physically large spiral
galaxies.  Contributing to the list are observations by
Baiesi-Pillastrini \markcite{Baie83}(1983); Bicay \& Giovanelli
(\markcite{Bica86a}\markcite{Bica86b}1986a,b, \markcite{Bica87}1987);
Freudling, Haynes, \& Giovanelli \markcite{Freu88}(1988); Giovanelli \&
Haynes (\markcite{Giov85a}\markcite{Giov85b}1985a,b,
\markcite{Giov89}1989, \markcite{Giov93}1993); Giovanelli et al.
\markcite{Giov86}(1986); Gordon \& Gottesman \markcite{Gord81}(1981);
Haynes \& Giovanelli (\markcite{Hayn84}1984, \markcite{Hayn86}1986,
\markcite{Hayn91}1991); Haynes et al.  \markcite{Hayn88}(1988); Hewitt,
Haynes, \& Giovanelli \markcite{Hewi83}(1983); Huchtmeier \& Richter
\markcite{Huch89}(1989); Kyazumov \markcite{Kyaz84}(1984); Lewis
\markcite{Lewi85}(1985); Nilson \markcite{Nils73}(1973); Romanishin
\markcite{Roma83}(1983); Sandage \& Tammann \markcite{Sand81}(1981);
Scaglia \& Sancisi \markcite{Scag88}(1988); Schneider et al.
\markcite{Schn86}(1986); Scodeggio \& Gavazzi \markcite{Scod93}(1993);
Stavely-Smith \& Davies \markcite{Stav88}(1988); de Vaucouleurs, de
Vaucouleurs, \& Corwin \markcite{deVa76}(1976); Wegner, Haynes, \&
Giovanelli \markcite{Wegn93}(1993).  Table~\ref{tab_large50inh} lists
the 50 largest spiral galaxies assembled from the above sources.  An
index of physical size, the product of Doppler velocity and angular
size, was used to rank the galaxies.  The same galaxy reported in
several sources will appear several times in the list from which
Table~\ref{tab_large50inh} was culled.  In most cases, the several
entries for the same galaxy were not wildly separated in the ranked
list.  Table~\ref{tab_large50hom} lists the 50 largest spiral galaxies
in the RC3 (de Vaucouleurs et al. \markcite{deVa91}1991), a homogeneous
catalog.  (In Table~\ref{tab_large50hom}, the velocity is with respect
to the 3K background.)  Tables~\ref{tab_large50inh}
and~\ref{tab_large50hom} include only spirals with morphological type
between Sb and Sc (inclusive) to help ensure enough $\rm H\,I$ for
emission line mapping (if desired) and to avoid the ragged spiral
structure of Sd and later types.  (The analysis of
\S\S\ref{sec_flatrc}-\ref{sec_shape} does not apply to multiple-armed
or flocculent spirals.)  Sometimes the literature source only specified
that the object was a spiral (morphological type ``S''), usually
because the object was too distant to show enough structure to
determine a Hubble subtype.  Such cases were permitted in the observing
list because the source material may have been poor and a much better
CCD frame could reveal more structure.  In addition, the inclination of
the galaxy was restricted to be in the range $27^{\circ}$ to $68^{\rm
\circ}$, based on major and minor axis angular sizes, so that the
spiral structure would be clear and later kinematic analysis would be
possible.  Many of the galaxies in Tables~\ref{tab_large50inh}
and~\ref{tab_large50hom} were observed as part of the observing program
reported in this paper.

Grand design spirals of all sizes were sought in addition to the
physically large spirals.  Objects from Elmegreen et al.
\markcite{Elme87}(1987) with arm class 12 (``grand design'') were
chosen in addition to objects whose photos in atlases (Sandage \& Bedke
\markcite{Sand88}1988; Sandage \& Tammann\markcite{Sand81}1981; Wray
\markcite{Wray88}1988) showed clear two-armed grand design spiral
structure.\placetable{tab_large50inh}\placetable{tab_large50hom}

Table~\ref{tab_obslist} shows the final list of objects observed with
brief comments about their spiral structure.  Two-armed grand design
spirals (called ``2-arm g.d.'' in Table~\ref{tab_obslist}) may have
spurs or feathers that are not so prominent as to give the disk a
multiple arm appearance.  Spiral structure that is ``not regular'' may
have arms with wildly varying pitch, there may be more than one
pattern, or some other deviation from two-armed grand design
structure.  Galaxies marked ``faint'' lack enough signal in the
exposure to categorize the spiral structure.

Limited morphological information that is useful to the discussion in
\S\ref{sec_discussion} is also reported in Table~\ref{tab_obslist}.
The presence of a bar is denoted by the symbol ``B''; an inner ring is
denoted by ``r'' and an inner pseudoring by ``p''; an outer ring is
denoted ``R'' and an outer pseudoring by ``P''.  (I apologize to Ron
Buta for this abbreviated classification scheme.)  If the galaxy is
not barred, it is designated ``S''.  When the feature is not
clear, the symbol is followed by a ``:''.  Deprojection (which
elongates round features) may cause uncertainty about the presence of a
bar.  The presence of a bar is also uncertain when the galaxy has an
S-shape:  then it is not clear if the nearly straight central segment
is a bar or merely the convergence of two open spiral arms on the
nucleus.  Inner rings may be round or pointy (that is,
lozenge-shaped).  Pseudorings are ring-like features that appear to be
formed by spiral arms, and may or may not be closed.  (See Buta
\markcite{Buta86}1986 on the morphology of rings and pseudorings.)
Inner pseudorings usually have a lozenge shape and are generally not
closed.  The distinction between an inner pseudoring and a pointy inner
ring is not clear.  For this paper, if the spiral arms followed
continuously from the ring segments, or if the ring segments did not
close, then the feature was called a pseudoring.  If the ring segments
were closed and the spiral arms were not obviously extensions of the
segments, then the feature was called an inner ring.  Outer pseudorings
have a fat dumbbell or figure-eight shape.\placetable{tab_obslist}

\section{Analysis}\label{sec_analysis}

A downhill simplex algorithm (Press et al.  \markcite{Pres88}1988)
fitted ellipses to isophotes by minimizing the isophotal deviations
around the ellipses.  The apparent shape of the disk was characterized
by the outer isophotes, which then specified the deprojection of the
image.  Deprojection was done with IRAF (Image Reduction and Analysis
Facility).  Table~\ref{tab_obslist} lists the position angles and
inclinations (to the nearest degree) used for the deprojection.
Indeterminate position angles result when the inclination is small.
The table lists formal errors on position angle and inclination that
were computed from the fitting errors of ellipses that characterize the
outer isophotes.  The fitting errors were empirically derived by
varying the starting point in parameter space for the downhill simplex
algorithm and compiling the deviations among the ending points.

The radial extent of spiral structure was measured on deprojected
images of the galaxies.  The inner and outer extent of the spiral
structure was marked conservatively, given the caveats in
\S\S\ref{sec_shape}--\ref{sec_caveats}.  (Inner spiral structure had to
be clear; outer spiral structure had to have reasonable
signal-to-noise, described below.)  The inner and outer radii of spiral
structure for the two-armed grand design spirals are listed in
Table~\ref{tab_spextent}.  Some galaxies show two patterns:  they have
faint spiral structure in their outer disks that is of a different
character than the bright inner arms (see Elmegreen \& Elmegreen
\markcite{Elme95}1995 on this topic).  For galaxies with an inner and
an outer pattern, separate computations were made for the end of the
bright spiral pattern and the end of its faint extension.

The column in Table~\ref{tab_spextent} labeled ``S/N'' indicates the
signal-to-noise ratio for the outer spiral structure detection.  Signal
was measured as a $7\times 7$-pixel ($5''\times 5''$, a typical spiral
arm width for galaxies observed here) average within the spiral arm at
the furthest extent of the spiral structure.  Local sky was determined
from an average of four such $7\times 7$-pixel averages in nearby,
almost emission-free areas.  The noise was determined as the average
standard deviation from the mean in the four sky measurements.

Surface photometry in the literature was used to calibrate the
instrumental surface brightness profiles of some galaxies.  The effect
of foreground stars on the measured profiles was reduced by ignoring
the brightest 10\% of the pixels in each radial zone.  Ignoring bright
pixels is admittedly not as accurate as fitting stellar profiles and
subtracting stars from each frame.  The object of this exercise is not
to derive accurate surface photometry, however, but rather to measure
the photometric depth of the data frames.

The calibrated surface brightness profiles are shown in
Figure~\ref{fig_surphot}.  There is excellent agreement in detail with
most profiles from the literature.  Small discrepancies in limited
radial ranges can be attributed to the effects of foreground stars that
were not uniformly treated in all published surface brightness
studies.  Surface brightness profiles are typically reliable to about
$27\rm\,mag\,arcsec^{-2}$.  At this isophotal level, the
signal-to-noise ratio (S/N) has dropped to 2.5, counting signal within
a typical seeing disk (about $2\rm\,arcsec$).\placefigure{fig_surphot}

The surface brightness at which the outer extent of spiral structure
was determined was typically much brighter than
$27\rm\,mag\,arcsec^{-2}$.  Using the calibrated surface photometry
shown in Fig.~\ref{fig_surphot} and outer spiral extents from
Table~\ref{tab_spextent}, the mean surface brightness at the outer
extent of the spiral structure was computed
(Table~\ref{tab_spbright}).  Spiral arms are significantly brighter
than the mean surface brightness, though.  Given that the gain of the
Tek1K chip (the most frequently used detector for this study) is
$7.4\rm\,e^-\,pix^{-1}$, its read noise ($6\rm\,e^-$) is insignificant
compared to sky noise for 30-minute exposures, and that the mean sky
brightness at the U.~S.  Naval Observatory Flagstaff Station at
$V$-band is $21.5\rm\,mag\,arcsec^{-2}$, the S/N per pixel equals unity
for the $26\rm\,mag\,arcsec^{-2}$ isophote.  Probably the brightness of
the spiral arm at the outer spiral extent reported in
Table~\ref{tab_spextent} can be, at faintest, about
$26\rm\,mag\,arcsec^{-2}$ at $V$-band.  Note that the surface
brightness at the outer extent of the spiral structure can be much
brighter if, for instance, only the extent of the bright, inner arms is
reported in Table~\ref{tab_spextent} (e.g., for NGC0521).

\begin{figure}
\plotone{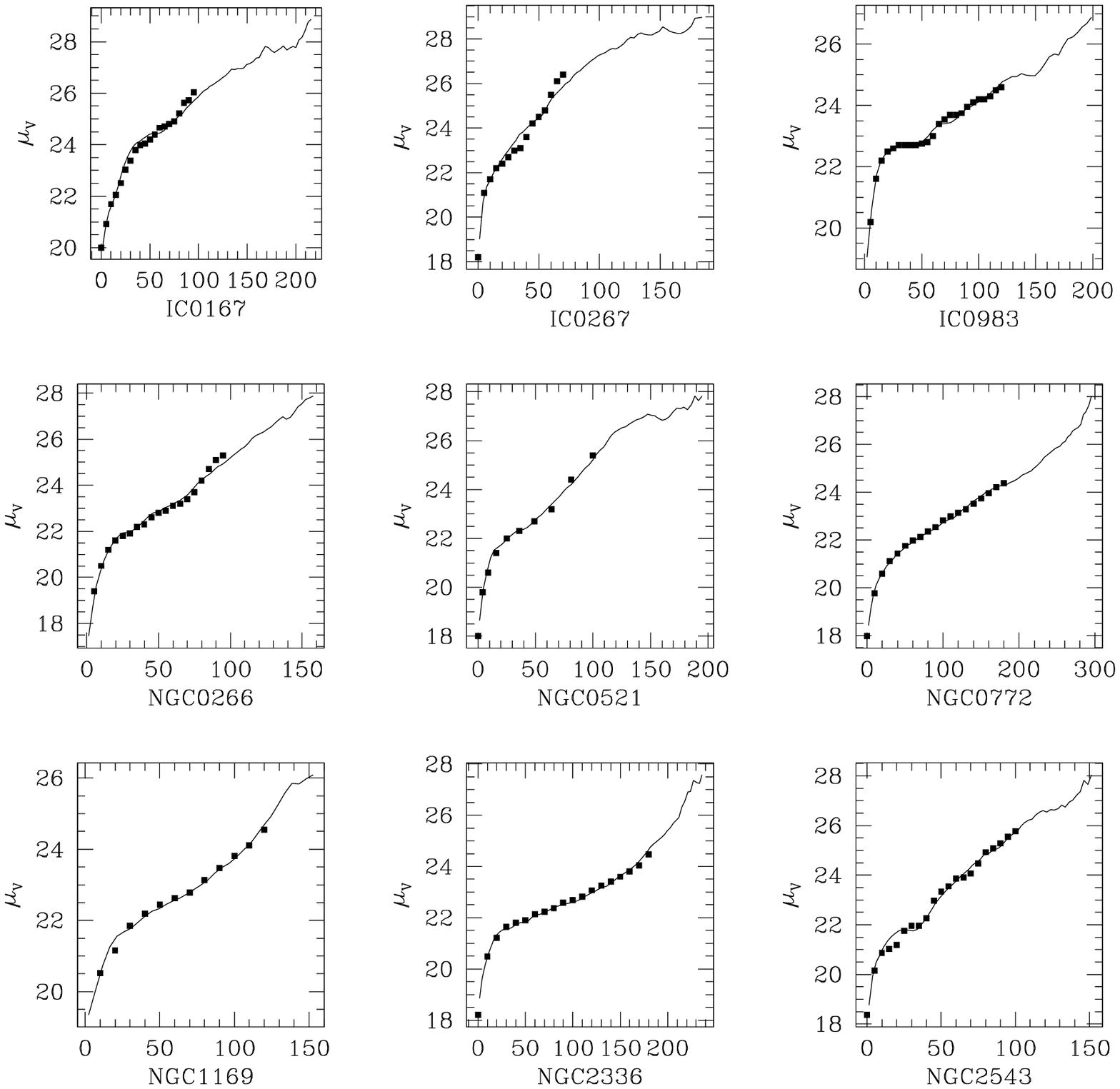}
\end{figure}

\begin{figure}
\figurenum{4}
\plotone{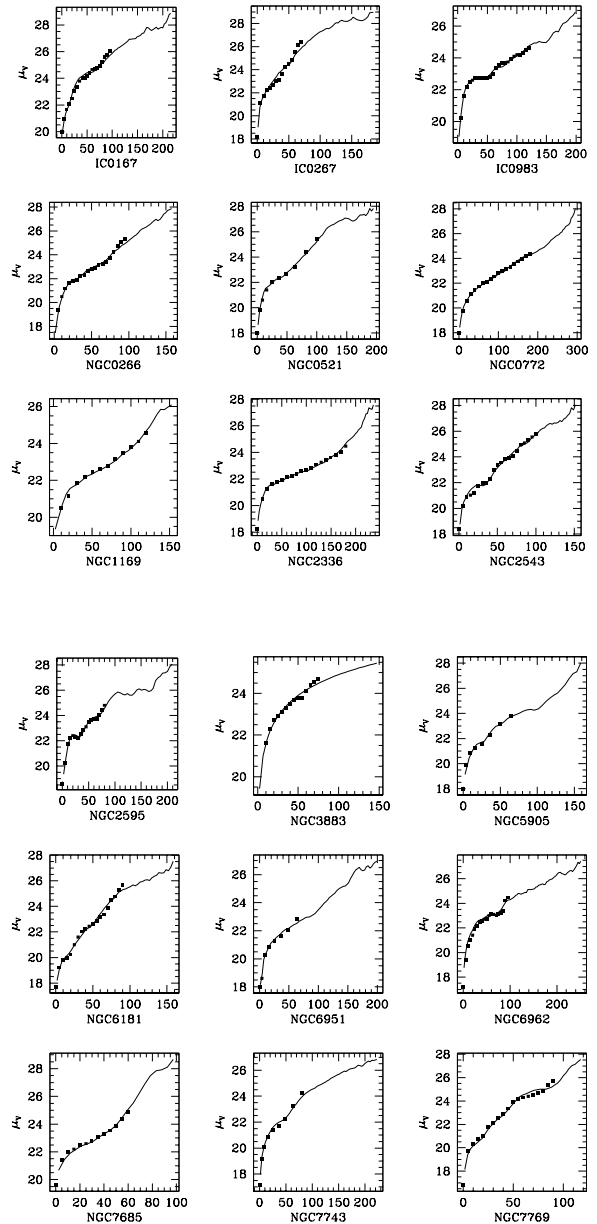}
\caption{These are calibrated surface brightness
profiles for some galaxies in the survey.  Surface brightness in
$V\rm\ mag\,arcsec^{-2}$ is graphed versus radius in arcseconds.  The
filled squares are data from the literature.  The lines are the
calibrated surface brightness profiles from CCD imaging observations
reported here, which are reliable to about
$27\rm\,V\,mag\,arcsec^{-2}.$\label{fig_surphot}}
\end{figure}

Given the generally large values of S/N at the outer spiral extent
listed in Table~\ref{tab_spextent} and the fact that the data frames
are rather deep (as shown by the surface photometry in
Fig.~\ref{fig_surphot}), nearly all the optically luminous spiral
structure is recorded on the data frames and generally it is recorded
comfortably above the noise threshold of the frames.

The results of checking for violations of equation~\ref{eq_ilrcr} and
adherence to equation~\ref{eq_ilrolr} are shown in
Table~\ref{tab_spextent} for the grand design spirals.  Violations are
assessed according to the ``conservative'' tests (\S\ref{sec_shape}).
A ``dot'' in the ``Pass?'' column indicates that the conservative test
is passed with a measured value of the ratio less than 1.71; a
``$\sim$'' in the ``Pass?'' column indicates a value of the ratio from
1.71 to 2.0; and if the measured ratio is over 2.0, there is an ``N''
in the ``Pass?'' column, indicating a violation of the conservative
ratio test.\placetable{tab_spextent}

\section{Discussion}\label{sec_discussion}

\begin{figure}
\figurenum{5}
\plotone{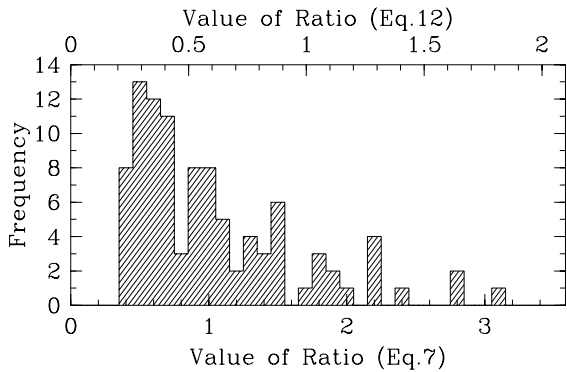}
\caption{Here are distributions of the
equation~\protect\ref{eq_ilrcr} (lower scale) and
equation~\protect\ref{eq_ilrolr} (upper scale) ratios for qualified
galaxies from Table~\protect\ref{tab_spextent}.\label{fig_histo}}
\end{figure}

The distribution of values of the equation~\ref{eq_ilrcr} and
equation~\ref{eq_ilrolr} ratios is shown in Figure~\ref{fig_histo}.
Only a single increment is shown for each galaxy in
Table~\ref{tab_spextent}, except that NGC5248 and NGC6632 have two
contributions because these galaxies have two distinct spiral
patterns.  In other cases where there are two entries for the same
galaxy in Table~\ref{tab_spextent}, the entry that represented the most
extensive, continuous, two-armed spiral structure was chosen.  Ratios
corresponding to the extent of extra or detached arms are not
illustrated in Fig.~\ref{fig_histo}.\placefigure{fig_histo}

Most of the galaxies have values of the equation~\ref{eq_ilrolr} ratio
that are less than unity.  This observation is consistent with spiral
density wave theory (spiral structure bounded by the Lindblad
resonances) and the fact that most rotation curves are approximately
flat.  Effects due to the shape of the rotation curve
(\S\ref{sec_shape}) can explain the few ratios larger than unity.

Surprisingly many grand design spirals have small values of the
equation~\ref{eq_ilrolr} ratio (peaked around 0.3 in
Fig.~\ref{fig_histo}).  Can theories of the creation of grand design
spiral structure explain this preponderance of low values?  Swing
amplification (see, e.g., Toomre \markcite{Toom81}1981) of disturbances
created by companions (Byrd \& Howard \markcite{Byrd92}1992) or a
groove (Sellwood \& Kahn \markcite{Sell91}1991) should create spirals
that reach both their ILR and OLR.  If the swing-amplified spiral is a
superposition of spiral density waves with a range of pattern speeds
(\S\ref{sec_caveats}), then measurement of the extent of the spiral
could lead to an equation~\ref{eq_ilrolr} ratio greater than unity, not
less.  But under modal theory (see, e.g., Bertin et al.
\markcite{Bert89a}\markcite{Bert89b}1989a,b), the innermost extent of
the spiral would be the $Q$-barrier, which is exterior to the ILR, thus
decreasing the equation~\ref{eq_ilrolr} ratio.

The best explanation for the low values of the equation~\ref{eq_ilrolr}
ratio involves the resonant interaction of spirals with bars and rings,
as will be discussed below.  There is a peak in the distribution of
values of equation~\ref{eq_ilrcr} near 0.5 (see Fig.~\ref{fig_histo}).
Most of the galaxies contributing to the peak are barred and ringed
galaxies.

\begin{figure}
\figurenum{6}
\plotone{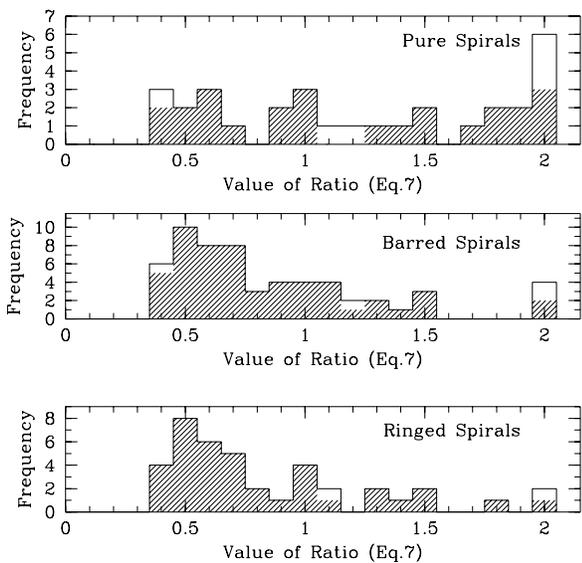}
\caption{The distribution of the
equation~\protect\ref{eq_ilrcr} ratios are graphed here separately for
pure spirals, barred spirals, and ringed spirals.  Uncertain
morphologies are graphed as an unshaded increment in the two
appropriate panels.  Barred spirals with rings contribute to two
panels.\label{fig_type}}
\end{figure}

The distributions of the values of the equation~\ref{eq_ilrcr} ratio
are graphed separately in Figure~\ref{fig_type} for three morphological
classes:  pure spirals with no sign of ring or bar, barred galaxies
(which may have a ring), and ringed galaxies (which may also have a
bar) including galaxies with pseudorings.  Galaxies showing signs of
strong interaction, such as an anomalously long spiral arm and a
probable companion, were omitted from all the distributions.  For
example, galaxies resembling Arp~82 or Arp~248, which have long and
obvious tidal tails, were excluded from the distributions.  Galaxies
resembling M51, which show a close interaction but no tidal tails, were
excluded because the projection of a tidal tail nearly along the line
of sight could masquerade as a spiral arm in such cases.  Galaxies
resembling M81 were included in the distributions.  Although M81 has
spiral structure that resulted from an interaction with a companion,
its spiral structure is a spiral density wave in the disk and there is
no sign of a tidal tail or other structure out of the plane of the
disk.  The galaxies in Table~\ref{tab_spextent} that were excluded from
the distributions are Anon0958--14, ESO437--G044, IC0167, IC1551,
IC2421, NGC0280, NGC2535, NGC4321, NGC7753, NGC7769, UGC04457,
UGC11453, and UGC11695.

As in Fig.~\ref{fig_histo}, when two ratio values are listed in
Table~\ref{tab_spextent}, the entry corresponding to the main two-armed
spiral structure is included in the Fig.~\ref{fig_type} histogram.  In
a few cases where the uncertain morphology could move a galaxy from the
barred or ringed class to the pure spiral class (for instance ``B:''
might be ``B'' or ``S'', and ``r:'' might be ``r'' or ``S''), the
galaxy is graphed on histograms for both possible classes as an
unshaded box.\placefigure{fig_type}

The values of the equation~\ref{eq_ilrcr} ratio are much more evenly
distributed for pure spirals than they are for the other morphological
classes.  (Pure spirals comprise less than one-third of the sample,
making their statistics somewhat noisier than those for barred or
ringed galaxies.)  The distributions of the equation~\ref{eq_ilrcr}
ratio for barred and ringed galaxies, in contrast to pure spirals, are
decidedly peaked at small values (around 0.5 to 0.7).  The different
appearance of the distributions in Fig.~\ref{fig_type} can be
quantified by a Kolmogorov-Smirnov test (cf. Press et
al.\markcite{Pres88}).  The probability that the ratios for pure
spirals are drawn from the same population as those for the barred
spirals is less than 1\%; the probability is just over 1\% for pure and
ringed spirals.  In contrast, the ratios for barred and ringed spirals
have a 97\% probability of being drawn from the same population.  The
last statistic is not surprising, since many galaxies have both bar and
ring features.  (Over half the barred galaxies in this study are also
ringed.)

A correlation between limited spiral extent and the presence of a bar
can be explained if, in many barred spirals, the bar and spiral are
resonantly related.  Modeling of barred spirals indicates that strong
bars end at or just short of their CR (Contopoulos
\markcite{Cont80}1980; cf. also Elmegreen \markcite{Elme96}1996).  If
the spiral patterns in strongly barred spirals had the same pattern
speed as the bar, then spirals could extend only between their CR and
OLR.  For such a barred spiral, the equation~\ref{eq_ilrolr} ratio
would be $(1+\sqrt{2}/2)/(2+\sqrt{2})=0.5$.

All bars are not strong, though.  The relationship, if any, between the
end of a weak bar and its resonances is not known.  Furthermore, it is
not established that bars and spirals are resonantly related when they
appear together in the same galaxy.  So the rationale explaining why
galaxies with bars tend to have less extensive spiral structure may be
only a partial explanation.  The explanation may be true often enough,
however, to explain why there is a peak around 0.5--0.7 in
Fig.~\ref{fig_type} for barred spirals but no peak for pure spirals.

A correlation between limited spiral extent and the presence of an
inner ring can also be explained if, as is fairly well established,
inner rings are resonance phenomena.  Inner rings and pseudorings seem
to be located at the inner 4:1 resonance (cf. Combes
\markcite{Comb96}1996), which is exterior to the ILR.  Limiting spiral
structure to the range between inner 4:1 and OLR also decreases the
equation~\ref{eq_ilrolr} ratio.

Other aspects of rings are neutral to the discussion or are
problematic.  The location of an outer ring at the OLR (Buta \& Crocker
\markcite{Buta91}1991) does not place an extra limitation on spiral
extent, and so does nothing to decrease the equation~\ref{eq_ilrolr}
ratio.  Many barred spirals also have an inner ring circumscribing the
bar.  It is not clear if it is possible to reconcile the presence of a
strong bar (ending inside its CR) with a surrounding inner ring (at the
4:1 resonance), although a weak bar and inner ring might be possible.

\subsection{Particular galaxies}\label{sec_particular}

Descriptions of the spiral structure of two-armed grand design spiral
galaxies with the largest values of the resonance ratios follow below,
where the relative merit of each galaxy for kinematic or resonance
analysis is assessed.  Because the shape of the rotation curve is not
known, the inclusion of a galaxy in this list is no guarantee that the
CR for the two-arm spiral pattern is within the optical disk (see
\S\ref{sec_shape}).

{\em NGC5829.}  This galaxy appears to be a member of a small group
(HCG73:  Hickson, Kindl, \& Auman \markcite{Hick89}1989).  One apparent
member of the group shows signs of recent interaction (tidal tail or
one-armed spiral emerging from an off-center, nucleated, disturbed or
irregular galaxy).  Spiral structure reaches a very small radius,
leading to a large equation~\ref{eq_ilrcr} ratio.  There is some danger
that the outer disk is distended or not planar owing to encounters
within the group.

{\em NGC6907.}  This galaxy may have a central oval, although the
morphology of the spiral arms may contribute to the illusion of the
oval.  In the inner disk, spiral structure is bright and sharply
defined, with regions of dust and star formation.  In the outer disk,
spiral arms are faint, smooth, and broad.  One spiral arm wraps through
considerably more phase than the other.

{\em IC0211.}  Spiral structure penetrates to very small radius in this
galaxy.  The arms are somewhat faint and diffuse, although they are
lightly dappled with regions of star formation.

{\em NGC1042.}  Much of the spiral structure of this galaxy is faint,
diffuse, and punctuated by bright knots (thus resembling IC0211).  The
spiral structure in the inner disk consists of two short, bright arcs.

{\em UGC10104.}  The inner disk has very regular, two-armed spiral
structure.  There is a radial gap at the end of one bright inner arm,
followed by a fainter, more diffuse, somewhat fragmentary outer arm.
The other bright inner arm shows similar behavior, except that the
corresponding outer arm tracks the outer rim of the inner arm for about
30 degrees in azimuth before the inner arm fades.  It is possible that
there are two resonantly related two-armed spiral patterns in this
galaxy.  If so, then the ratio reported in Table~\ref{tab_spextent} is
inappropriately large.

{\em NGC0132.}  This galaxy may have a central oval, or else (like
NGC6907) the inner spiral arms mimic a lozenge shape.  One arm has a
ragged and feathery appearance.

{\em NGC0173.}  The spiral structure in the inner disk has a feathery
appearance and it is somewhat difficult to trace the arms continuously
into the outer disk.  The spiral arms in the outer disk are narrow and
dotted by regions of star formation.

{\em NGC7685.}  The spiral structure is somewhat feathery and is
complicated by the emergence of a short arm from the interarm region
between the two main spiral arms.

The following galaxies marginally pass the conservative
test (indicated by a ``$\sim$'' in Table~\ref{tab_spextent}).

{\em UGC04643.}  The spiral structure is very regular and winds through
a turn and a half.  The arms make an abrupt transition from bright to
faint after about a full turn.  The spiral arms in the outer disk are
still rather narrow and are not diffuse, in contrast to other galaxies
with different arm brightnesses in the inner and outer disk.

{\em UGC09808.}  The spiral arms are exceptionally thin.  There is some
irregularity in the emergence of the two-armed spiral.  It appears that
there is a single spiral arm in the inner disk that leads continuously,
except for a kink or cusp, to one of the two main outer arms.  The
second main arm emerges in a manner symmetrical to the first, but
without obvious connection to the inner disk.  A short arm or arc
emerges from the interarm region interior to the main arm with the
kink.  The short arm is initially bright but quickly dims and
broadens.  Both of the main arms make a transition from narrow and
bright to fainter and more diffuse in the outer disk.

{\em NGC3897.}  The spiral arms are narrow and emerge from nearby the
nucleus.  The otherwise regular morphology is complicated by one arm
forking in the outer disk.  The inner tine of the forked arm winds a
half turn into the outer disk as a broader, faint arm.  The outer tine
of the forked arm is initially the brighter tine but it fades after a
quarter turn.

{\em IC2627.}  This galaxy has two bright arms in the inner disk that
abruptly and symmetrically transition to broader, faint arms in
the outer disk.  One of the main arms is considerably longer than the
other.  A very faint, third arm that winds half a turn in the outer
disk emerges from an interarm region between the two main inner arms.

{\em NGC4079.} The spiral structure is very regular and winds inward to
a rather small radius.  There is an abrupt change from bright, narrow
arms in the inner disk to broad, faint arms in the outer disk,
suggesting a resonance crossing.  This galaxy is an exemplary case for
kinematic and resonance analysis.  The author has obtained VLA $\rm
H\,I$ data with which to map its velocity field.

\section{Conclusions}\label{sec_concl}

This paper describes an observational project that examined 140 spiral
galaxies, measuring the visual extent of their spiral structure.  The
purpose of the project was to find galaxies for which the corotation
resonance of the two-armed grand design spiral structure was likely to
be within the bright optical disk.  The likelihood that the corotation
resonance is within the optical disk can be assessed by comparing the
measured ratio of outer to inner spiral extent with the value of the
canonical ratio (equation~\ref{eq_ilrcr}).

The test using the canonical ratio is not conclusive because of
variations due to rotation curve shape.  It is also possible that there
is no inner Lindblad resonance.  To counter our ignorance of the
rotation curve, a test with a more conservative value of the ratio than
equation~\ref{eq_ilrcr} should be used (here, an extra factor of two).

A few galaxies were found that have large spiral extent and whose
corotation resonance radius is very likely to be within the bright
optical disk.  Among the best candidates for kinematic study are
NGC5829, NGC6907, IC0211, NGC1042, and NGC0132.

It was incidentally found that barred and ringed spirals statistically
have more limited spiral extent than pure spirals.  This observation
can be explained if many bars end at their inner 4:1 or corotation
resonance and if the bar and spiral are part of the same pattern.  The
location of inner rings at the inner 4:1 resonance also limits spiral
extent, but the location of outer rings at the outer Lindblad resonance
places no extra constraint on spiral extent.

\acknowledgements
This research was supported in part by the Director's Research Fund at
the Space Telescope Science Institute.  I thank Otto Richter for an
electronic copy of his table of $\rm H\,I$ observations and Riccardo
Giovanelli for electronic copies of several tables summarizing $\rm
H\,I$ observations.  I have benefited from conversations with Ken
Freeman, Piet van der Kruit, and Alar Toomre.  Arne Henden kindly
observed and provided processed CCD frames of six of the galaxies for
this project.  Thanks to the referee, Bruce Elmegreen, for his comments
and criticisms that led to improvements of this paper.  Provisional
images of some galaxies were obtained during a preliminary phase of
this research using the Guide Stars Selection System Astrometric
Support Program developed at the Space Telescope Science Institute
(STScI is operated by the Association of Universities for Research in
Astronomy, Inc. for NASA).  This research has made use of the NASA/IPAC
Extragalactic Database (NED), which is operated by the Jet Propulsion
Laboratory, Caltech, under contract with the National Aeronautics and
Space Administration.  Electronic copies of several tables of galaxy
and redshift information were obtained from the National Space Science
Data Center and the Astronomical Data Center.

\newpage
\begin{deluxetable}{rllrrc}
\tablewidth{0pt}
\tablecaption{Largest 50 Galaxies (Various Sources)\label{tab_large50inh}}
\tablehead{\colhead{$a\times V$}&\colhead{Name}&\colhead{Alt. Name}&
\colhead{$a$}&\colhead{$V$}&\colhead{Ref.}\\
&&&\colhead{arcmin}&\colhead{$\rm km\,s^{-1}$}}
\startdata
61875&Malin1&&2.50&24750&(1)\nl
40709&UGC02885&&7.02&5799&(1)\nl
36210&UGC08058&Anon1254+57&1.7&21300&(2)\nl
34330&IC1551&UGC00268&2.6&13204&(3)\nl
31292&IC0983&UGC09061&5.75&5442&(1)\nl
30697&UGC11473&&6.60&4651&(1)\nl
29476&UGC10104&Anon1555+30&2.95&9992&(4)\nl
26472&UGC02339&&1.60&16545&(12)\nl
25571&NGC3883&UGC06754&3.64&7025&(1)\nl
25341&NGC5174&UGC08475&3.70&6849&(1)\nl
25254&NGC5688&&9.00&2806&(1)\nl
25056&IC0226&UGC01922&2.30&10894&(1)\nl
24488&NGC2206&UGCA123&3.90&6279&(1)\nl
23870&UGC05813&UGC05813&1.8&13261&(3)\nl
22631&UGC01886&&4.50&5029&(11)\nl
22523&NGC2713&UGC04691&5.75&3917&(1)\nl
22464&UGC09515&&1.6&14040&(5)\nl
22198&NGC1241&&5.50&4036&(1)\nl
22128&UGC02151&&2.0&11064&(8)\nl
21755&UGC12128&&1.1&19777&(10)\nl
21725&NGC0772&UGC01466&8.86&2452&(1)\nl
21577&NGC1961&UGC03334&5.50&3923&(1)\nl
21122&IC1142&UGC10055&1.5&14081&(5)\nl
21040&NGC1085&UGC02241&3.1&6787&(6)\nl
20619&NGC5981&UGC09948&4.10&5029&(1)\nl
20544&NGC2446&UGC04027&2.10&9783&(1)\nl
20504&IC4381&UGC09073&2.20&9320&(1)\nl
20477&UGC00520&&1.00&20477&(12)\nl
20222&NGC4939&&6.50&3111&(1)\nl
20122&NGC5619&UGC09255&2.4&8384&(7)\nl
20075&NGC0497&UGC00915&2.5&8030&(2)\nl
19958&UGC12084&&1.6&12474&(9)\nl
19839&UGC09025&&1.10&18035&(1)\nl
19497&Anon2233+34&&2.95&6609&(1)\nl
19345&NGC6007&UGC11079&1.82&10629&(4)\nl
19180&NGC1530&UGC03013&7.80&2459&(1)\nl
19166&NGC2336&UGC03809&8.70&2203&(1)\nl
19164&IC0213&UGC01719&2.30&8332&(12)\nl
19089&IC1173&UGC10180&1.83&10431&(1)\nl
18985&NGC1169&UGC02503&7.95&2388&(1)\nl
18946&NGC5172&UGC08477&4.70&4031&(1)\nl
18921&NGC7753&UGC12780&3.5&5406&(8)\nl
18795&UGC04457&Anon0829+19A&1.70&11056&(4)\nl
18717&UGC10405&&1.7&11010&(3)\nl
18634&NGC5905&UGC09797&5.50&3388&(1)\nl
18600&NGC0858&&1.51&12318&(4)\nl
18591&UGC01212&&1.70&10936&(11)\nl
18522&NGC3968&UGC06895&2.90&6387&(1)\nl
18403&NGC0536&UGC01013&3.55&5184&(1)\nl
18389&UGC09107&&1.2&15324&(5)\nl
\enddata
\tablecomments{(1) Huchtmeier \& Richter\markcite{Huch89} (2)
Nilson\markcite{Nils73} (3) Haynes \& Giovanelli \markcite{Hayn84}(1984) (4)
de Vaucouleurs et al. \markcite{deVa76}(1976) (5) Freudling et
al.\markcite{Freu88} (6) Kyazumov\markcite{Kyaz84} (7) Haynes \& Giovanelli
\markcite{Hayn91}(1991) (8) Giovanelli \& Haynes \markcite{Giov85a}(1985a)
(9) Giovanelli et al. \markcite{Giov86}(1986) (10) Giovanelli \& Haynes
\markcite{Giov89}(1989) (11) Wegner et al.\markcite{Wegn93} (12) Giovanelli
\& Haynes \markcite{Giov93}(1993)}
\end{deluxetable}

\begin{deluxetable}{rllrr}
\tablewidth{0pt}
\tablecaption{Largest 50 Galaxies (RC3)\label{tab_large50hom}}
\tablehead{\colhead{$a\times V$}&\colhead{Name}&\colhead{Alt. Name}&
\colhead{$a$}&\colhead{$V_{\rm 3K}$}\\
&&&\colhead{arcmin}&\colhead{$\rm km\,s^{-1}$}}
\startdata
32641&IC1551&UGC00268&2.57&12699\nl
30509&IC0983&UGC09061&5.37&5681\nl
28442&UGC02885&&5.01&5675\nl
27559&UGC04219&&2.19&12597\nl
26732&UGC10104&Anon1555+30&2.69&9932\nl
23874&IC0226&UGC01922&2.24&10664\nl
23306&UGC03139&&3.72&6273\nl
21677&NGC3883&UGC06754&2.95&7345\nl
21156&UGC09234&&1.91&11103\nl
20745&UGC04457&Anon0829+19A&1.82&11400\nl
20550&UGC09515&&1.45&14217\nl
20300&NGC1085&UGC02241&3.09&6569\nl
20084&UGC12128&&1.05&19180\nl
19614&UGC03531&&1.62&12094\nl
19601&NGC1961&UGC03334&5.01&3911\nl
19300&NGC4939&&5.62&3432\nl
19054&UGC02151&&1.78&10715\nl
18854&NGC5619&UGC09255&2.19&8618\nl
18588&UGC03374&&3.02&6155\nl
18392&NGC4501&UGC07675&7.08&2598\nl
18285&UGC09025&&1.00&18285\nl
18148&UGC10250&&0.93&19446\nl
18117&NGC3968&UGC06895&2.69&6731\nl
18114&UGC01886&&3.89&4656\nl
18113&NGC6007&UGC10079&1.70&10666\nl
18078&NGC3314B&&3.80&4755\nl
17738&UGC09107&&1.15&15449\nl
17481&NGC1365&&11.22&1558\nl
17156&Anon0958$-$14&&1.82&9428\nl
16912&NGC0309&&3.16&5348\nl
16847&PGC07941&&2.63&6405\nl
16804&UGC01810&Anon0218+39A&2.29&7335\nl
16772&NGC7753&UGC12780&3.47&4837\nl
16740&UGC03252&&2.75&6078\nl
16620&UGCA228&Anon1108$-$09&2.04&8140\nl
16598&NGC0772&UGC01466&7.59&2188\nl
16546&NGC5514&UGC09102&2.19&7563\nl
16395&IC3165&UGC07384&1.91&8604\nl
16279&NGC2336&UGC03809&7.41&2196\nl
16273&NGC6744&&20.42&797\nl
16191&NGC4535&UGC07727&7.08&2287\nl
16084&UGC02351&&1.95&8249\nl
16020&UGC08510&&1.10&14610\nl
15884&UGC04207&&1.66&9571\nl
15865&NGC6177&UGC10428&1.70&9342\nl
15783&NGC5652&UGC09334&2.04&7730\nl
15760&NGC6632&UGC11226&3.39&4651\nl
15692&IC1222&UGC10461&1.70&9240\nl
15599&NGC5230&UGC08573&2.19&7130\nl
15445&NGC6674&UGC11308&4.68&3302\nl
\enddata
\end{deluxetable}

\begin{deluxetable}{llrrrrllc}
\tablewidth{0pt}
\tablecaption{Galaxies Observed\label{tab_obslist}}
\tablehead{\colhead{Name}&\colhead{Alt. Name}&\colhead{PA}&\colhead{err.}&
\colhead{Incl.}&\colhead{err.}&\colhead{Description}&\colhead{Type}&
\colhead{Notes}\\
&&\colhead{deg.}&\colhead{deg.}&\colhead{deg.}&\colhead{deg.}}
\startdata
Anon0958$-$14&&\ldots&\ldots&\ldots&\ldots&2-arm g.d.&Bp&(1)\nl
ESO437$-$G044&AM1039$-$283&\ldots&\ldots&0&\ldots&2-arm g.d.&S\nl
IC0167&UGC01313&86&3&49&2&2-arm g.d.&B&(1,3)\nl
IC0211&UGC01678&39&3&46&2&2-arm g.d.&Bp\nl
IC0267&UGC02368&14&7&37&3&2-arm g.d.&B\nl
IC0421&UGCA111&\ldots&\ldots&0&\ldots&mixed&Bp\nl
IC0983&UGC09061&\ldots&\ldots&0&\ldots&multiple arms&Bp\nl
IC1142&UGC10055&148&10&32&7&multiple arms&Bp\nl
IC1237&UGC10621&20&1&71&1&2-arm g.d.&B:p:&(4)\nl
IC1302&&43&3&59&2&2-arm g.d.&S\nl
IC1516&UGC12852&45&17&11&6&multiple arms&Sr\nl
IC1525&UGC12883&26&5&30&6&2-arm g.d.&Bp\nl
IC1551&UGC00268&\ldots&\ldots&\ldots&\ldots&two tidal arms&S&(2)\nl
IC2421&UGC04658&61&31&22&12&2-arm g.d.&S&(3)\nl
IC2627&UGCA227&134&15&34&4&2-arm g.d.&S\nl
IC4381&UGC09073&126&6&50&5&2-arm g.d.&S&(3)\nl
NGC0036&UGC00106&9&1&61&2&2-arm g.d.&Bp\nl
NGC0132&UGC00301&46&3&39&2&2-arm g.d.&B:\nl
NGC0173&UGC00369&94&5&37&5&2-arm g.d.&S\nl
NGC0180&UGC00380&158&4&32&3&mixed&Bp\nl
NGC0266&UGC00508&118&8&17&5&2-arm g.d.&Bp:\nl
NGC0280&UGC00534&104&5&51&4&2-arm g.d.&S\nl
NGC0497&UGC00915&131&4&58&2&multiple arms&Bp\nl
NGC0521&UGC00962&\ldots&\ldots&0&\ldots&mixed&Bp:\nl
NGC0536&UGC01013&63&2&70&2&2-arm g.d.&Br\nl
NGC0562&UGC01049&6&26&18&10&2-arm g.d.&S\nl
NGC0606&UGC01126&109&6&30&9&3-arm g.d.&Bp\nl
NGC0753&UGC01437&121&7&46&0&multiple arms&S\nl
NGC0772&UGC01466&120&4&47&1&not regular&S\nl
NGC0841&UGC01676&131&1&61&1&2-arm g.d.&P:Sr\nl
NGC1042&&182&11&20&8&2-arm g.d.&S\nl
NGC1169&UGC02503&34&4&46&2&multiple arms&Br\nl
NGC1376&&\ldots&\ldots&0&1&multiple arms&S\nl
NGC1417&&6&2&62&1&2-arm g.d.&B:p\nl
NGC1784&&111&6&35&5&multiple arms&Br\nl
NGC2206&UGCA123&134&3&54&2&2-arm g.d.&Sr\nl
NGC2280&UGCA131&158&4&65&2&2-arm g.d.&S&(1,4)\nl
NGC2336&UGC03809&182&2&52&1&multiple arms&Sp\nl
NGC2460&UGC04097&17&3&45&3&mixed&S\nl
NGC2535&UGC04264&\ldots&\ldots&\ldots&\ldots&2-arm g.d.&Sp&(2)\nl
NGC2543&UGC04273&42&4&51&2&2-arm g.d.&B\nl
NGC2595&UGC04422&41&7&42&1&2-arm g.d.&Bp\nl
NGC2608&UGC04484&43&3&38&1&multiple arms&Bp:\nl
NGC2713&UGC04691&107&2&66&1&2-arm g.d.&Br\nl
NGC2718&UGC04707&88&21&19&16&2-arm g.d.&Br\nl
NGC2935&UGCA169&167&3&31&3&2-arm g.d.&B\nl
NGC3038&&121&2&56&2&mixed&S\nl
NGC3054&UGCA187&120&3&49&2&2-arm g.d.&B\nl
NGC3183&UGC05582&155&7&40&10&2-arm g.d.&B\nl
NGC3319&UGC05789&39&2&55&2&2-arm g.d.&B\nl
NGC3347&&175&3&64&3&2-arm g.d.&Bp&(3)\nl
NGC3381&UGC05909&75&20&17&16&2-arm g.d.&B\nl
NGC3433&UGC05981&46&8&37&2&2-arm g.d.&S\nl
NGC3478&UGC06069&131&2&60&2&multiple arms&B\nl
NGC3513&UGCA224&92&5&36&6&2-arm g.d.&Bp\nl
NGC3577&UGC06257&\ldots&\ldots&2&2&2-arm g.d.&Bp\nl
NGC3583&UGC06263&124&4&38&1&2-arm g.d.&B\nl
NGC3642&UGC06385&35&35&5&5&2-arm g.d.&S\nl
NGC3883&UGC06754&160&12&34&4&mixed&B\nl
NGC3897&UGC06784&92&2&35&4&2-arm g.d.&Sp\nl
NGC3963&UGC06884&\ldots&\ldots&0&\ldots&2-arm g.d.&B\nl
NGC3968&UGC06895&7&3&48&3&multiple arms&B\nl
NGC3992&UGC06937&69&1&53&1&multiple arms&Bp\nl
NGC4079&UGC07067&127&2&46&4&2-arm g.d.&S\nl
NGC4321&UGC07450&27&8&33&1&2-arm g.d.&B\nl
NGC4662&UGC07917&137&10&13&7&multiple arms&Bp\nl
NGC5172&UGC08477&95&1&50&1&one-armed&S\nl
NGC5230&UGC08573&\ldots&\ldots&2&2&multiple arms&Br\nl
NGC5248&UGC08616&99&3&36&2&2-arm g.d.&S\nl
NGC5409&UGC08938&41&3&46&5&2-arm g.d.&Sr\nl
NGC5619&UGC09255&6&1&62&3&2-arm g.d.&Bp\nl
NGC5829&UGC09673&55&6&32&5&2-arm g.d.&S\nl
NGC5905&UGC09797&128&5&34&3&2-arm g.d.&Bp\nl
NGC6177&UGC10428&23&3&43&3&2-arm g.d.&BPr\nl
NGC6181&UGC10439&2&1&61&2&2-arm g.d.&B\nl
NGC6412&UGC10897&49&28&7&7&2-arm g.d.&B\nl
NGC6632&UGC11226&150&2&59&1&two patterns?&Sp\nl
NGC6674&UGC11308&137&3&50&1&multiple arms&Br\nl
NGC6907&UGCA418&97&35&17&15&2-arm g.d.&B:\nl
NGC6951&UGC11604&150&6&31&3&2-arm g.d.&Bp\nl
NGC6956&UGC11619&122&32&16&8&2-arm g.d.&Br\nl
NGC6962&UGC11628&77&1&38&2&2-arm g.d.&B\nl
NGC7042&UGC11702&128&3&35&2&2-arm g.d.&S\nl
NGC7309&&138&66&12&12&3-arm g.d.&S\nl
NGC7407&UGC12230&151&1&61&2&2-arm g.d.&S\nl
NGC7685&UGC12638&176&14&40&5&2-arm g.d.&Sr\nl
NGC7738&UGC12757&70&7&52&2&2-arm g.d.&BP\nl
NGC7743&UGC12759&114&9&29&4&2-arm g.d.&BP\nl
NGC7753&UGC12780&33&3&47&3&2-arm g.d.&B\nl
NGC7769&UGC12808&25&7&25&11&mixed&S\nl
NGC7782&UGC12834&178&2&54&2&faint&S\nl
NGC7819&UGC00026&105&3&43&4&2-arm g.d.&B\nl
UGC00250&&125&1&48&3&multiple arms&Br\nl
UGC00520&&138&1&60&1&2-arm g.d.&S\nl
UGC00850&&65&4&58&2&2-arm g.d.&B:r&(3)\nl
UGC01919&&35&10&33&4&2-arm g.d.&Br\nl
UGC02151&&109&1&59&1&three arms&S\nl
UGC02507&&5&2&55&1&2-arm g.d.&B\nl
UGC02582&&10&9&22&14&multiple arms&S\nl
UGC02885&&44&1&65&1&multiple arms&B:\nl
UGC03120&&115&2&41&2&2- or 3-arm g.d.&S\nl
UGC03252&&40&3&41&2&mixed&S\nl
UGC03294&&125&2&61&5&multiple arms&Br\nl
UGC03325&&34&6&32&0&2-arm g.d.&B:\nl
UGC03374&&33&24&16&14&2-arm g.d.&Br\nl
UGC03375&&49&1&61&1&multiple arms&B:p\nl
UGC03531&&139&2&48&3&2-arm g.d.&Br\nl
UGC03593&&46&1&58&2&2-arm g.d.&S\nl
UGC04207&&21&6&56&2&2-arm g.d.&Sr\nl
UGC04219&&122&3&42&4&2-arm g.d.&S\nl
UGC04457&Anon0829+19A&171&9&52&3&2-arm g.d.&S\nl
UGC04643&&94&5&36&4&2-arm g.d.&S\nl
UGC04706&&130&10&33&9&2-arm g.d.&B\nl
UGC04884&&41&4&47&3&2-arm g.d.&S\nl
UGC05813&&40&3&52&3&multiple arms&Br\nl
UGC06093&&\ldots&\ldots&1&1&2-arm g.d.&Br&(1)\nl
UGC07065&&144&1&35&2&2-arm g.d.&BPr&(1,3)\nl
UGC08058&Anon1254+57&\ldots&\ldots&\ldots&\ldots&2-arm g.d.&S&(2)\nl
UGC08510&&\ldots&\ldots&9&9&multiple arms&S\nl
UGC09025&&64&3&53&3&multiple arms&B:\nl
UGC09107&&150&6&45&4&2-arm g.d.&Sr&(3)\nl
UGC09808&&\ldots&\ldots&0&\ldots&2-arm g.d.&S\nl
UGC10104&Anon1555+30&19&36&9&9&2-arm g.d.&S\nl
UGC10405&&40&9&37&3&multiple arms&B\nl
UGC10837&&151&1&59&2&2-arm g.d.&Bp:\nl
UGC11406&&7&3&53&4&2-arm g.d.&B\nl
UGC11453&Anon1930+54&79&1&40&1&2-arm g.d.&S\nl
UGC11473&&128&4&62&2&not regular&S\nl
UGC11585&&82&2&48&3&2-arm g.d.&Br\nl
UGC11695&Anon2109--01&107&4&56&4&2-arm g.d.&S\nl
UGC11809&&31&1&61&0&not regular&B\nl
UGC11919&Anon2206+40&23&6&44&4&2-arm g.d.&S\nl
UGC12084&&96&14&30&7&2-arm g.d.&Br\nl
UGC12128&&130&1&61&0&2-arm g.d.&S\nl
UGC12164&&121&3&52&2&2-arm g.d.&S\nl
UGC12199&&95&4&55&1&2-arm g.d.&Br\nl
UGC12646&&48&1&23&7&2-arm g.d.&BPr\nl
UGC12776&&95&12&37&4&2-arm g.d.&Br\nl
UGC12792&&56&10&42&2&2-arm g.d.&BR\nl
UGCA228&Anon1108$-$09&10&51&8&8&multiple arms&Bp\nl
\enddata
\tablecomments{``g.d.'' means ``grand design''; ``mixed'' means having
both 2-armed grand design and multiple arm morphology in different
parts of the disk; (1) exposure is insufficient to show reliable outer
isophotes; (2) merger, interacting, or tidal, not deprojected; (3)
spiral structure influences shape of outer isophotes; (4) deprojected
according to inner disk isophotes.}
\end{deluxetable}

\begin{deluxetable}{llrrllccrl}
\tablewidth{0pt}
\tablecaption{Spiral Extent\label{tab_spextent}}
\tablehead{\colhead{Name}&\colhead{Alt. Name}&\colhead{$R_1$}&\colhead{$R_2$}&
\colhead{Eq.~\ref{eq_ilrcr}}&\colhead{Pass?}&\colhead{Eq.~\ref{eq_ilrolr}}&
\colhead{Pass?}&\colhead{S/N}&\colhead{Notes}\\
&&\colhead{arcsec}&\colhead{arcsec}}
\startdata
Anon0958$-$14&&7&51&2.06&N&1.21&.&3.0&(1)\nl
ESO437$-$G044&AM1039$-$283&25&112&1.31&.&0.77&.&1.6&(1)\nl
IC0167&UGC01313&13&104&2.41&N&1.36&.&6.8&(10)\nl
IC0211&UGC01678&7&65&2.78&N&1.57&.&14.0&\nl
IC0267&UGC02368&56&83&0.43&.&0.24&.&5.8&\nl
IC0421&UGCA111&21&73&1.02&.&0.57&.&9.3&(7,8)\nl
IC1237&UGC10621&13&69&1.55&.&0.87&.&10.9&(4)\nl
IC1302&&9&28&0.94&.&0.53&.&11.6&\nl
IC1525&UGC12883&14&51&1.05&.&0.59&.&14.0&\nl
IC1551&UGC00268&16&77&1.39&.&0.81&.&3.0&(1,10)\nl
IC2421&UGC04658&22&63&0.84&.&0.49&.&3.0&(10)\nl
IC2627&UGCA227&14&85&1.79&$\sim$&1.05&.&5.5&(4)\nl
&&14&105&2.21&N&1.30&.&3.0&(3,12)\nl
IC4381&UGC09073&12&61&1.47&.&0.86&.&2.4&(4,5)\nl
NGC0036&UGC00369&18&85&1.37&.&0.80&.&4.0&(2)\nl
NGC0132&UGC00301&8&56&2.19&N&1.28&.&2.8&(14)\nl
NGC0173&UGC00369&15&110&2.18&N&1.23&.&3.1&\nl
NGC0180&UGC00380&15&30&0.59&.&0.33&.&39.7&(7,8)\nl
NGC0266&UGC00508&34&108&0.94&.&0.55&.&5.3&\nl
NGC0280&UGC00534&11&42&1.09&.&0.64&.&23.9&\nl
NGC0521&UGC00962&19&44&0.68&.&0.40&.&14.9&(2,7,8)\nl
NGC0536&UGC01013&63&100&0.47&.&0.27&.&11.1&\nl
NGC0562&UGC01049&6&33&1.68&.&0.95&.&16.9&(4)\nl
NGC0841&UGC01676&25&45&0.52&.&0.30&.&5.5&\nl
NGC1042&&12&94&2.38&N&1.39&.&8.1&\nl
NGC1417&&15&73&1.46&.&0.85&.&8.3&\nl
NGC2206&UGCA123&20&67&0.99&.&0.58&.&12.3&\nl
NGC2280&UGCA131&22&72&0.99&.&0.58&.&66.9&(2,8)\nl
&&22&243&3.31&N&1.94&$\sim$&4.4&(3,7)\nl
NGC2460&UGC04097&82&160&0.57&.&0.34&.&4.7&(5,9)\nl
NGC2535&UGC04264&20&160&2.41&N&1.41&.&4.2&(1)\nl
NGC2543&UGC04273&23&90&1.15&.&0.65&.&3.5&(3)\nl
NGC2595&UGC04422&33&88&0.77&.&0.45&.&4.8&\nl
NGC2713&UGC04691&45&74&0.49&.&0.28&.&24.7&\nl
NGC2718&UGC04707&32&48&0.44&.&0.26&.&11.6&\nl
NGC2935&UGCA169&43&130&0.89&.&0.50&.&8.9&(2)\nl
&&43&155&1.06&.&0.60&.&1.2&(3)\nl
NGC3038&&34&78&0.67&.&0.38&.&14.0&(9,16)\nl
NGC3054&UGCA187&19&63&0.99&.&0.58&.&23.5&(7,8)\nl
&&19&98&1.54&.&0.90&.&8.3&(3,12)\nl
NGC3183&UGC05582&27&61&0.67&.&0.39&.&18.6&\nl
NGC3319&UGC05789&56&170&0.89&.&0.52&.&12.8&\nl
NGC3347&&74&131&0.52&.&0.30&.&67.5&\nl
NGC3381&UGC05909&14&43&0.90&.&0.53&.&15.3&\nl
NGC3433&UGC05981&21&49&0.68&.&0.40&.&31.8&(7,8)\nl
NGC3513&UGCA224&22&93&1.25&.&0.70&.&13.3&\nl
NGC3577&UGC06257&12&52&1.29&.&0.72&.&1.2&\nl
NGC3583&UGC06263&27&68&0.75&.&0.42&.&32.6&\nl
NGC3642&UGC06385&28&128&1.35&.&0.76&.&11.3&(3)\nl
NGC3883&UGC06754&18&40&0.63&.&0.37&.&8.2&(7,8)\nl
NGC3897&UGC06784&12&71&1.82&$\sim$&1.06&.&3.1&\nl
NGC3963&UGC06884&18&66&1.08&.&0.61&.&12.8&\nl
NGC4079&UGC07067&12&75&1.76&$\sim$&1.03&.&5.0&\nl
NGC4321&UGC07450&60&155&0.76&.&0.44&.&39.7&\nl
NGC5248&UGC08616&28&100&1.06&.&0.62&.&103.9&(8)\nl
&&100&219&0.64&.&0.38&.&3.8&(5)\nl
NGC5409&UGC08938&21&41&0.58&.&0.34&.&10.1&\nl
NGC5619&UGC09255&27&59&0.64&.&0.38&.&21.2&\nl
NGC5829&UGC09673&5&52&3.10&N&1.75&$\sim$&14.4&\nl
NGC5905&UGC09797&27&105&1.14&.&0.67&.&5.4&(2)\nl
&&27&128&1.40&.&0.82&.&4.7&(3,12)\nl
NGC6177&UGC10428&23&44&0.56&.&0.33&.&14.1&\nl
NGC6181&UGC10439&22&59&0.79&.&0.44&.&10.4&(4)\nl
NGC6412&UGC10897&10&53&1.52&.&0.85&.&27.2&\nl
NGC6632&UGC11226&27&42&0.46&.&0.27&.&33.1&(8)\nl
&&55&88&0.47&.&0.27&.&16.4&(5)\nl
NGC6907&UGCA418&11&107&2.85&N&1.67&.&5.1&\nl
NGC6951&UGC11604&54&107&0.59&.&0.34&.&3.7&\nl
NGC6956&UGC11619&20&46&0.68&.&0.40&.&15.1&\nl
NGC6962&UGC11628&42&73&0.51&.&0.30&.&15.1&\nl
NGC7042&UGC11702&10&49&1.40&.&0.82&.&25.1&\nl
NGC7407&UGC12230&10&30&0.88&.&0.52&.&21.2&(2)\nl
NGC7685&UGC12638&10&67&2.05&N&1.16&.&5.2&(4)\nl
NGC7738&UGC12757&42&65&0.46&.&0.27&.&3.3&\nl
NGC7743&UGC12759&39&55&0.42&.&0.24&.&20.9&\nl
NGC7753&UGC12780&23&101&1.31&.&0.77&.&8.0&\nl
NGC7769&UGC12808&28&94&1.00&.&0.58&.&4.2&(9,10,16)\nl
NGC7819&UGC00026&21&52&0.74&.&0.43&.&7.6&\nl
UGC00520&&10&17&0.50&.&0.29&.&7.3&\nl
UGC00850&&7&31&1.30&.&0.76&.&12.9&\nl
UGC01919&&19&35&0.54&.&0.32&.&5.9&\nl
UGC02507&&14&27&0.56&.&0.33&.&9.0&(2)\nl
&&14&32&0.66&.&0.39&.&4.3&(3,11)\nl
UGC03252&&13&27&0.63&.&0.37&.&68.7&(7,8)\nl
UGC03325&&9&38&1.19&.&0.70&.&4.9&\nl
UGC03374&&41&71&0.50&.&0.28&.&4.0&(4)\nl
UGC03531&&15&45&0.88&.&0.52&.&5.5&\nl
UGC03593&&11&34&0.91&.&0.51&.&25.5&\nl
UGC04207&&13&48&1.05&.&0.59&.&5.3&(2)\nl
UGC04219&&13&64&1.49&.&0.87&.&1.6&\nl
UGC04457&&7&66&2.80&N&1.64&.&3.5&(10)\nl
UGC04643&&10&63&1.89&$\sim$&1.07&.&5.2&\nl
UGC04706&&23&40&0.52&.&0.30&.&4.6&\nl
UGC04884&&23&30&0.38&.&0.21&.&8.2&\nl
UGC06093&&8&36&1.32&.&0.77&.&2.7&\nl
UGC07065&&26&54&0.61&.&0.34&.&6.6&\nl
UGC09107&&17&42&0.74&.&0.43&.&4.8&\nl
UGC09808&&7&45&1.87&$\sim$&1.05&.&6.2&(4)\nl
UGC10104&Anon1555+30&10&75&2.20&N&1.24&.&5.0&\nl
UGC10837&&25&54&0.65&.&0.38&.&5.1&\nl
UGC11406&Anon1906+42&46&71&0.45&.&0.25&.&19.7&\nl
UGC11453&Anon1930+54&23&57&0.73&.&0.41&.&19.4&(10)\nl
UGC11585&&15&34&0.66&.&0.39&.&15.2&\nl
UGC11695&Anon2109--01&10&28&0.84&.&0.49&.&5.8&(4,13)\nl
&&10&56&1.70&.&1.00&.&2.4&(10,14)\nl
UGC11919&Anon2206+40&10&37&1.04&.&0.61&.&13.8&\nl
UGC12084&&19&35&0.54&.&0.32&.&4.4&\nl
UGC12128&&20&28&0.42&.&0.24&.&3.1&\nl
UGC12164&&17&58&1.02&.&0.60&.&6.1&\nl
UGC12199&&35&49&0.42&.&0.23&.&11.2&\nl
UGC12646&&31&44&0.42&.&0.25&.&6.3&\nl
UGC12776&&47&121&0.76&.&0.43&.&3.3&\nl
UGC12792&&11&22&0.61&.&0.35&.&11.7&\nl
\enddata
\tablecomments{In the ``Pass?'' columns, ``.'' means the ratio passes
the ``conservative'' ratio test (\S\ref{sec_shape}), ``$\sim$'' means
the ratio is marginally in violation (value between 1.71 and 2), and
``N'' means the ratio fails the conservative test.  (1) not
deprojected; (2) main pattern or bright arms only; (3) including faint
or broad arms in outer disk; (4) some irregularity, regular arms only;
(5) two patterns, outer pattern only; (6) includes both 2-arm patterns;
(7) outer disk has multiple arms; (8) inner 2-arm pattern only; (9)
inner disk has multiple arms; (10) possibly tidal or interacting; (11)
extension of main 2-arm pattern; (12) detached from main 2-arm pattern;
(13) shorter of two arms; (14) longer of two arms; (15) deprojection
may not be appropriate; (16) outer two-arm pattern.}
\end{deluxetable}

\begin{deluxetable}{llc}
\tablewidth{0pt}
\tablecaption{Spiral Arm Brightness at Outer Extent\label{tab_spbright}}
\tablehead{\colhead{Name}&\colhead{Alt. Name}&\colhead{$\mu_V$}\\
&&\colhead{$\rm\,mag\,arcsec^{-2}$}}
\startdata
IC0167&UGC01313&26.0\nl
IC0267&UGC02368&26.5\nl
NGC0266&UGC00508&25.6\nl
NGC0521&UGC00962&22.5\nl
NGC2543&UGC04273&25.2\nl
NGC2595&UGC04422&25.1\nl
NGC3883&UGC06754&23.6\nl
NGC5905&UGC09797&25.7\nl
NGC6181&UGC10439&23.3\nl
NGC6951&UGC11604&23.8\nl
NGC6962&UGC11628&23.2\nl
NGC7685&UGC12638&25.9\nl
NGC7743&UGC12759&22.6\nl
NGC7769&UGC12808&25.5\nl
\enddata
\end{deluxetable}

\end{document}